\documentclass{article}
\usepackage{spconf,amsmath,graphicx}
\usepackage{multirow}
\usepackage{comment}
\usepackage{amsmath}
\usepackage{amssymb}
\usepackage{amsfonts}
\usepackage{enumerate}
\usepackage{siunitx}
\usepackage{booktabs}
\usepackage{bbding}
\usepackage{color}

\newlength\savedwidth

\newcommand{\argmax}{\mathop{\rm arg~max}\limits} 
\usepackage{bm}
\newcommand{\bmit}[1]{{\mbox{\boldmath $#1$}}}

\title{Onoma-to-wave: Environmental sound synthesis\\from onomatopoeic words}

\name{Yuki Okamoto$^{1}$,
       Keisuke Imoto$^{2}$,
       Shinnosuke Takamichi$^{3}$, 
       }
 \secondlinename{	  
       Ryosuke Yamanishi$^{4}$, 
       Takahiro Fukumori$^{1}$, 
       Yoichi Yamashita$^{1}$
       }

 \address{$^1$ Ritsumeikan University, Japan $^2$ Doshisha University, Japan\\$^3$ The University of Tokyo, Japan $^4$ Kansai University, Japan
  }

\begin{document}
\maketitle
\begin{abstract}
In this paper, we propose a framework for environmental sound synthesis from onomatopoeic words.
As one way of expressing an environmental sound, we can use an onomatopoeic word, which is a character sequence for phonetically imitating a sound.
An onomatopoeic word is effective for describing diverse sound features. 
Therefore, using onomatopoeic words for environmental sound synthesis will enable us to generate diverse environmental sounds.
To generate diverse sounds, we propose a method based on a sequence-to-sequence framework for synthesizing environmental sounds from onomatopoeic words.
We also propose a method of environmental sound synthesis using onomatopoeic words and sound event labels.
The use of sound event labels in addition to onomatopoeic words enables us to capture each sound event’s feature depending on the input sound event label.
Our subjective experiments show that our proposed methods achieve higher diversity and naturalness than conventional methods using sound event labels. 
\end{abstract}
\section{Introduction}
\label{sec:intro}
Environmental sound synthesis is a research field of sound generation and is the task of generating natural environmental sounds.
Many environmental sounds are used in the production of movies, games, and other contents \cite{Lloyd_ACMI3DGG_01}.
However, there is a limit to the amount of environmental sound data that is openly available.
In addition, there are cases where the environmental sound that exactly matches the required sound does not exist. 
Therefore, it is possible to solve these problems by using environmental sound synthesis.
Moreover, environmental sound synthesis has great potential for many applications such as supporting movie and game production \cite{Lloyd_ACMI3DGG_01, Kong_ICASSP2019_01, Wang_VR2017_01, Zhou_CVPR2018_01}, and data augmentation for sound event detection and scene classification \cite{Salamon_WASPAA2017_01, Gontier_ICASSP_2020}.

In recent years, some methods of environmental sound synthesis using deep learning approaches have been developed \cite{Kong_ICASSP2019_01, okamoto_arXiv_2019, Liu_arXiv_2020}.
One of the methods of environmental sound synthesis uses sound event labels as the input \cite{okamoto_arXiv_2019}.
The method enables the generation of environmental sounds expressing sound events.
In this method, since only sound event labels are input to the system, similar sounds are generated for the given sound event class; thus, the generated sounds are not sufficiently varied.
Another possibility of environmental sound synthesis is to use onomatopoeic words, which are character sequences that phonetically imitate sounds.
According to the studies of Lemaitre and Rocchesso \cite{Lemaitre_2018} and Sundaram and Narayanan \cite{Sundaram_2006}, onomatopoeic words are effective for expressing the features of audio samples. 
For example, when expressing {\it the sound of a whistle} using onomatopoeic words, we can distinguish the sounds with different durations and pitches using the length of the phoneme sequence, such as ``py u'' (short whistle) and ``p i i i'' (long whistle).
Based on the idea of mapping onomatopoeic words to environmental sounds, Kawai developed KanaWave \cite{KanaWave}, software that generates environmental sounds from onomatopoeic words.
KanaWave generates environmental sounds by simply connecting multiple sounds corresponding to the input onomatopoeic words, each of which is associated with a specific sound in a one-to-one correspondence.
Therefore, the sounds generated by KanaWave do not have sufficient naturalness and diversity. 
To utilize environmental sounds in media content, such as in animation and movie production, an environmental sound synthesis method that can generate synthesized sounds with high naturalness and diversity is required.

In this paper, we propose environmental sound synthesis from onomatopoeic words using a statistical approach.
Statistical methods make it possible to automatically learn the correspondence between environmental sounds and onomatopoeic words from large amounts of data with high diversity.
Even if there is a large dataset, the diversity of generated sounds is limited because the conventional method generates sounds by combining sounds in a dataset.
On the other hand, a statistical method enables us to generate more diverse synthesized sounds than conventional methods.
In the proposed method, we utilize the sequence-to-sequence conversion framework (seq2seq framework) \cite{seq2seq_arXiv_2014} to generate environmental sounds from onomatopoeic words. 
The seq2seq framework is often used in sequence-to-sequence conversions, such as those in speech synthesis and neural machine translation, and has shown high performance in many studies \cite{Tacotron_arXiv_2017,Ikawa_ICASSP2018_01, Drossos_2017}.
The seq2seq framework uses several layers of recurrent neural network (RNN), which can model time-series information.
Therefore, the seq2seq framework enables us to generate environmental sounds by considering the phoneme sequence for an onomatopoeic word.
We also propose a method of environmental sound synthesis using sound event labels, which are used in the conventional method, and onomatopoeic words.
The use of onomatopoeic words and sound event labels enables us to capture each sound event’s feature depending on the input sound event label.

The remainder of this paper is structured as follows. 
In Sec.~\ref{sec:proposed}, we describe the proposed methods of environmental sound synthesis from an onomatopoeic word. 
In Sec.~\ref{sec:experiments}, we report subjective experiments carried out to evaluate the performance of environmental sound synthesis from an onomatopoeic word.
Finally, we summarize and conclude this paper in Sec.~\ref{sec:conclusion}. 
\begin{figure}[t!]
\centering
\begin{center}
\includegraphics[scale=0.60]{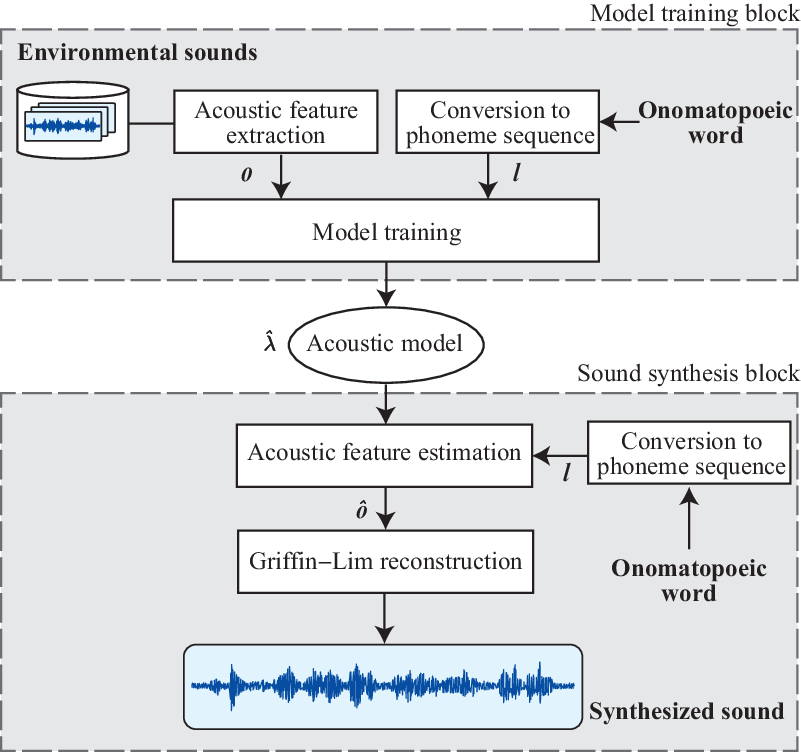}
\caption{Overview of environmental sound synthesis using onomatopoeia}
\label{fig:propsed_method}
\end{center}
\end{figure}

\section{Proposed Method}
\label{sec:proposed}
\subsection{Overview of Environmental Sound Synthesis from Onomatopoeic Words}
\label{subsec:overview_proposed_method}
Figure \ref{fig:propsed_method} shows the framework of environmental sound synthesis from onomatopoeic words.
This approach consists of a model training block and a sound synthesis block.
In the model training block, acoustic feature sequence ${\bmit o}$ and phoneme sequence ${\bmit l}$ are extracted from environmental sounds and onomatopoeic words, respectively.
Acoustic model parameter ${\bm \lambda}$ is estimated using extracted features ${\bmit o}$ and ${\bmit l}$ as follows:
\begin{equation}
\label{eq:train}
\hat{{\bm \lambda}} = \argmax_{{\bm \lambda}} P({\bmit o} \mid {\bmit l},{\bm \lambda}).
\end{equation}
In this paper, we propose two model training methods as follows.
\begin{enumerate}[(I)]
\item Model training method using only onomatopoeic words as input to network (Sec.~\ref{sec:proposed}-B-1)\\[-5pt]
\item Model training method using onomatopoeic words and sound event labels as input to network (Sec.~\ref{sec:proposed}-B-2)
\end{enumerate}
We will detail the model training methods in Sec.~\ref{subsec:model_training_method}. 
In the sound synthesis block, phoneme sequence ${\bmit l}$ is converted from an input onomatopoeic word.
Acoustic feature sequence ${\bmit o}$ is estimated from a phoneme sequence ${\bmit l}$ of the onomatopoeic word and acoustic model $\hat{{\bm \lambda}}$ as follows:
\begin{equation}
\label{eq:synthesis}
\hat{{\bmit o}} = \argmax_{{\bmit o}} P({\bmit o} \mid {\bmit l},\hat{{\bm \lambda}}).
\end{equation}
Finally, we reconstruct an environmental sound wave from estimated acoustic feature sequence $\hat{{\bmit o}}$ using the Griffin--Lim algorithm \cite{Griffin_Lim}. 
\subsection{Proposed Model Training Methods}
\label{subsec:model_training_method}
\subsubsection{Environmental Sound Synthesis Using Onomatopoeic Words}
\label{subsubsec:proposed_onomatopoeia}
\begin{figure}[t!]
\centering
\begin{center}
\includegraphics[scale=0.50]{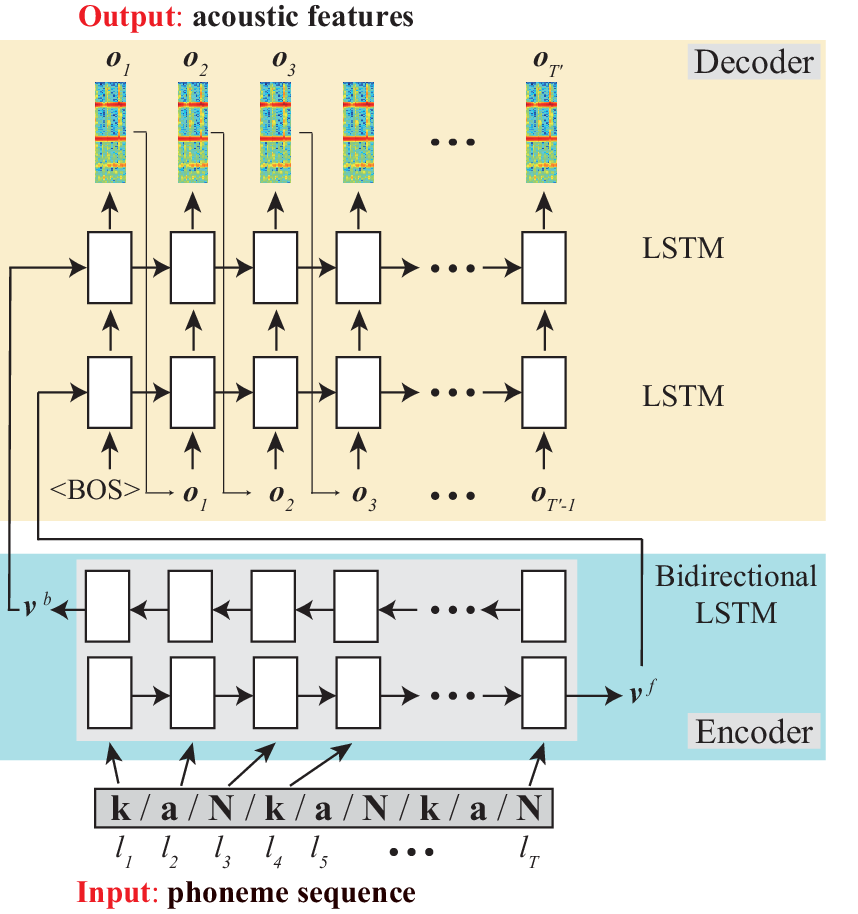}
\caption{Environmental sound synthesis from onomatopoeic words}
\label{fig:propsed_input_onomatopoeia}
\end{center}
\end{figure}

Figure \ref{fig:propsed_input_onomatopoeia} shows an overview of model training using onomatopoeic words.
To synthesize environmental sounds from onomatopoeic words, we employ the seq2seq framework \cite{seq2seq_arXiv_2014}. 
The seq2seq framework comprises an encoder and a decoder.
Our method uses one-layered bidirectional long short-term memory (BiLSTM) as the encoder and two-layered long short-term memory (LSTM) as the decoder. 
As shown in Fig.~\ref{fig:propsed_input_onomatopoeia}, a phoneme sequence of the onomatopoeic word, ${\bmit l} = \{l_1,...l_T\}$, is input to the encoder.
The encoder extracts feature vectors ${\bm \nu} = [{\bm \nu}^{f}, {\bm \nu}^{b}]$ from input sequence ${\bmit l}$.
Superscripts $f$ and $b$ indicate forward and backward networks, respectively.
In unidirectional LSTM, the beginning features tend to be lost when the sequence is long.
Therefore, using BiLSTM for the encoder, we can expect to extract a feature vector ${\bm \nu}$ that captures entire onomatopoeic words from past and future directions.
The decoder estimates acoustic feature sequence ${\bmit o} = \{{\bmit o}_1,...,{\bmit o}_{T^{'}}\}$ from extracted feature vectors ${\bm \nu}$ in the encoder as follows:

\begin{equation}
\label{eq:train_onomatope}
p({\bmit o}_1,...,{\bmit o}_{T'} \mid l_1,...,l_T) = \prod_{t=1}^{T'} p({\bmit o}_t \mid \bm{\nu}, {\bmit o}_1,...,{\bmit o}_{t-1}).
\end{equation}
Using two-layered LSTM for the decoder, we can expect to estimate acoustic features by considering features in the forward and backward directions of onomatopoeic words extracted by the encoder. 
The L1 norm between the estimated acoustic feature sequence ${\bmit o}$ and the target feature sequence at each time step is used as the loss function. 
\subsubsection{Environmental Sound Synthesis Using Onomatopoeic Words and Sound Event Labels}
\label{subsubsec:proposed_onomatopoeia_eventLabels}
The method of environmental sound synthesis using only onomatopoeic words is expected to enable the control of the time-frequency structural features of synthesized sounds, such as sound duration.
The method of environmental sound synthesis using only onomatopoeic words will generate diverse sounds.
However, for example, the onomatopoeic word ``p a N'' could be considered to fit multiple sound events, such as {\it the sound of shooting guns} and {\it balloons breaking}.
Therefore, we cannot control the frequency property associated with the sound categories using only onomatopoeic words.
To control the frequency characteristics of sound events, we utilize sound event labels in addition to onomatopoeic words.

\begin{figure}[t!]
\centering
\begin{center}
\includegraphics[scale=0.50]{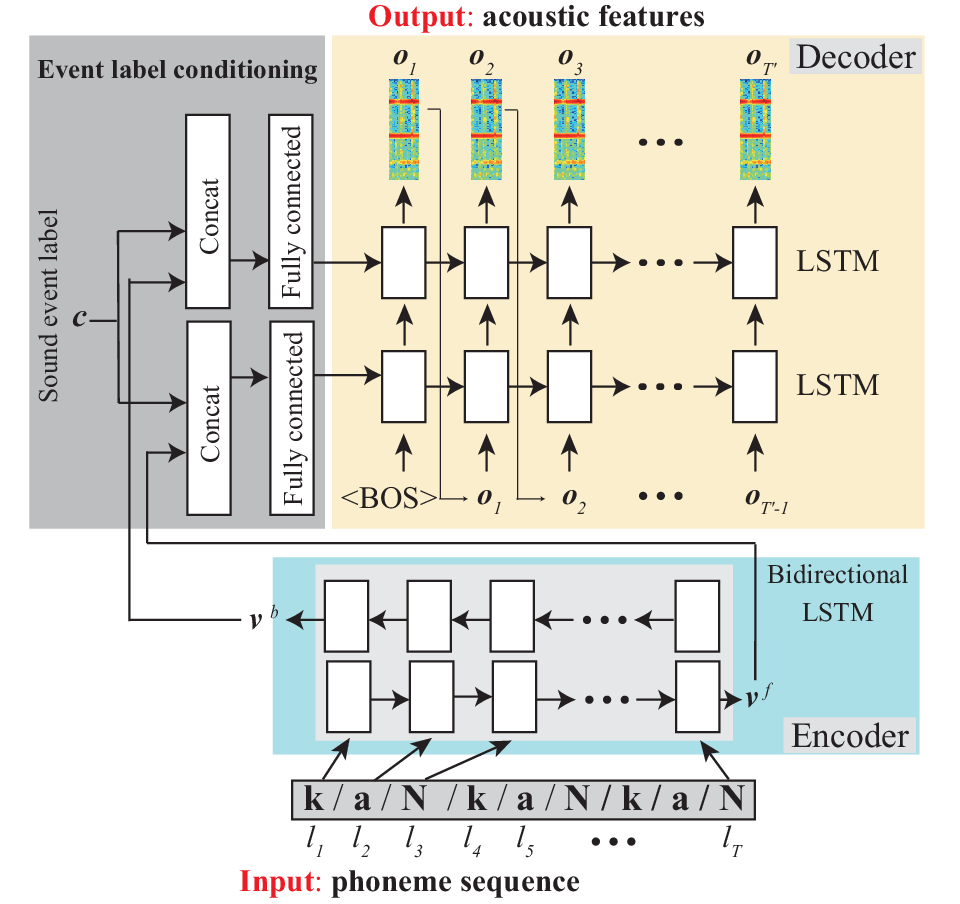}
\caption{Environmental sound synthesis from onomatopoeic words and sound event labels}
\label{fig:propsed_input_onomatopoeia+event_label}
\end{center}
\end{figure}

Figure \ref{fig:propsed_input_onomatopoeia+event_label} shows an overview of model training using onomatopoeic words and sound event labels.
The method uses the seq2seq framework comprising one-layered BiLSTM as the encoder and two-layered LSTM as the decoder. 
The seq2seq-based intersequence conversion may involve conditioning on the decoder to control the decoder's output features \cite{Jia_NIPS_2018, Ikawa_DCASE2019_01, cooper_ICASSP2020, park_arXiv_2019}. 
In the proposed method, sound event labels ${\bmit c}$ represented as one-hot vectors and extracted feature vectors ${\bm \nu}$ are concatenated and given as the initial state of the decoder.
The decoder estimates acoustic feature sequence ${\bmit o} = \{{\bmit o}_1,...,{\bmit o}_{T^{'}}\}$ from extracted feature vectors ${\bm \nu}$ in the encoder and sound event labels ${\bmit c}$ as follows: 

\begin{equation}
\label{eq:train_onomatope_EventLabel}
p({\bmit o}_1,...,{\bmit o}_{T'} \mid l_1,...,l_T) = \prod_{t=1}^{T'} p({\bmit o}_t \mid {\bm \nu}, {\bmit o}_1,...,{\bmit o}_{t-1}, {\bmit c}).
\end{equation}
The L1 norm between the estimated acoustic feature sequence ${\bmit o}$ and the target at each time step is used as the loss function. 
%
\section{Experiments}
\label{sec:experiments}
The synthesized sounds must have high naturalness and diversity to use synthesized sounds as background sounds or sound effects in movies or games.
From this viewpoint, we conducted two types of subjective test. 
For synthesized sounds, we evaluated their (I) naturalness and (II) sound diversity as environmental sounds.
We aim to achieve the same level of quality as natural sound in terms of both the naturalness and diversity of the generated sound.
\begin{table}[t!]
\vspace{-5pt}
\small
\caption{Experimental conditions}
\label{table:experiment}
\centering
\begin{tabular}{ll}
    \hline
    &\\[-8pt]
    Sound length & 1--2 s\\
    Sampling rate & 16,000 Hz\\
    Waveform encoding & 16-bit linear PCM\\
    \hline
    &\\[-8pt]
    Acoustic feature & log-amplitude spectrogram\\
    Window length for FFT & 0.128 s (2,048 samples) \\
    Window shift for FFT & 0.032 s (512 samples) \\
    \hline
    &\\[-8pt]
    Encoder LSTM layers & 1 \\
    \# units in encoder LSTM & 512 \\
    Decoder LSTM layers & 2 \\
    \# units in decoder LSTM & 512, 512 \\
    Event label dimensions & 10 \\
    Teacher forcing rate & 0.6\\
    Batch size & 5 \\
    Optimizer  & RAdam \cite{RAdam_ICLR_2020}\\
    \hline
\end{tabular}
\end{table}
%
\subsection{Experimental Conditions}
\label{subsec:experimental_conditions}
For the evaluation, we used 10 types of sound event ({\it bell ringing}, {\it alarm clock}, {\it manual coffee grinder}, {\it cup clinking}, {\it drum}, {\it maracas}, {\it electric shaver}, {\it tearing paper}, {\it trash box banging}, and {\it whistle}) contained in the Real World Computing Partnership-Sound Scene Database (RWCP-SSD) \cite{Nakamura_ASJ1999}. 
We used a total of 1,000 samples (100 samples $\times$ 10 sound events), in which 95 samples of each sound event were used for model training and the others were used for the subjective test. 
For the onomatopoeic words corresponding to each sound sample, we used the dataset in RWCP-SSD-Onomatopoeia \cite{okamoto_DCASE_2020}.
Each sound sample has more than 15 onomatopoeic words, and we used 15 onomatopoeic words per audio sample for model training for a total of 14,250 onomatopoeic words (15 onomatopoeic words $\times$ 950 audio samples).
Table~\ref{table:experiment} shows the experimental conditions and parameters used for the proposed methods. 
In this study, we use the log-amplitude spectrogram as an acoustic feature. 
The generated audio samples are available on our web page\footnote{https://y-okamoto1221.github.io/Onoma\_to\_wave\_Demonstration/}.

\subsection{Subjective Evaluations}
%
Following the evaluation perspective described at the beginning of Sec.~\ref{sec:experiments}, we conducted the following two sets of experiments:
\subsubsection{Experiment I: evaluation of naturalness for environmental sounds}
The target sound of this paper is a sound that is comfortable as an environmental sound and that expresses the input onomatopoeic word.
There are two perspectives of naturalness that should be satisfied.
For this reason, we designed several experiments to evaluate each perspective.
In Experiments I-1 and II-2, we presented environmental sounds and the onomatopoeic word used for the input, and evaluated how acceptable or expressive the presented sounds were in relation to the onomatopoeic word.  
In Experiment I-3, only the sound was presented to evaluate its naturalness as an environmental sound, and the sound itself was simply evaluated in terms of ``quality as an environmental sound.''
\\
\\
\begin{itemize}
        \item{\bf Experiment I-1: acceptance level of synthesized sounds for onomatopoeic words}\\
        We presented pairs of a sound (natural or synthesized) and an onomatopoeic word.
        The listener graded the acceptance level of the synthesized and natural sounds for onomatopoeic words on a scale of 1 (highly unacceptable) to 5 (highly acceptable).\\[-5pt]
        \item{\bf Experiment I-2: expressiveness of synthesized sounds for onomatopoeic words}\\
        We presented pairs of a sound (natural or synthesized) and an onomatopoeic word. 
        The listener graded the expressive level of the synthesized and natural sounds for onomatopoeic words on a scale of  1 (very unexpressive) to 5 (very expressive).\\[-5pt]
       \item{\bf Experiment I-3: naturalness of environmental sounds}\\
        We presented a natural or synthesized sound. 
        The listener graded the naturalness of the synthesized and natural sounds on a scale of 1 (very unnatural as an environmental sound) to 5 (very natural as an environmental sound).\\
\end{itemize}
\begin{table}[t!]
\vspace{-5pt}
\caption{Number of synthesized sounds used for subjective test}
\vspace{3pt}
\label{table:experimental_samples}
\centering
\small
\begin{tabular}{lcccc}
    \hline
    &\\[-10pt]
    \multirow{2}{*}{Experiment} & \multirow{2}{*}{\# labels} & \# samples & \multirow{2}{*}{\# listeners} & \# total\\[-1pt]
    & & in each label & & samples\\
    \hline
    &\\[-8pt]
    Exp. I-1 & 10 & 10 & 30 & 3,000\\
    Exp. I-2 & 10 & 10 & 30 & 3,000\\
    Exp. I-3 & 10 & 5 & 30 & 1,500\\
    Exp. II-1 & 5 & 5 & 30 & 750\\
    Exp. II-2 & 10 & 2-3 & 50 & 1,300\\
    \hline
\end{tabular}
\end{table}
\begin{table}[t!]
\vspace{-5pt}
\caption{List of synthesis methods evaluated for each evaluation metric}
\vspace{3pt}
\label{table:list_evaluation}
\centering
\footnotesize
\begin{tabular}{lccccc}
    \hline
    &\\[-8pt]
    Method & Exp. I-1 & Exp. I-2 & Exp. I-3 & Exp. II-1 & Exp. II-2\\
    &\\[-9pt]
    \hline
    &\\[-7pt]
    Natural sound & \Checkmark & \Checkmark & \Checkmark & & \\[3pt]
    WaveNet &  &  & \Checkmark & \Checkmark &\\[2pt]
    KanaWave & \Checkmark & \Checkmark & \Checkmark &  & \\[3pt]
    Seq2seq & \multirow{2}{*}{\Checkmark} & \multirow{2}{*}{\Checkmark} & \multirow{2}{*}{\Checkmark} &  & \multirow{2}{*}{\Checkmark} \\
    \bf{(}{\bf proposed}\bf{)} &&&&&\\[4pt]
    Seq2seq & \multirow{3}{*}{\Checkmark} & \multirow{3}{*}{\Checkmark} & \multirow{3}{*}{\Checkmark} & \multirow{3}{*}{\Checkmark} & \multirow{3}{*}{\Checkmark}\\
    + event label &&&&&\\
    \bf{(}{\bf proposed}\bf{)} &&&&&\\
    \hline
\end{tabular}
\end{table}
\begin{figure}[t!]
\centering
\begin{center}
\includegraphics[scale=0.45]{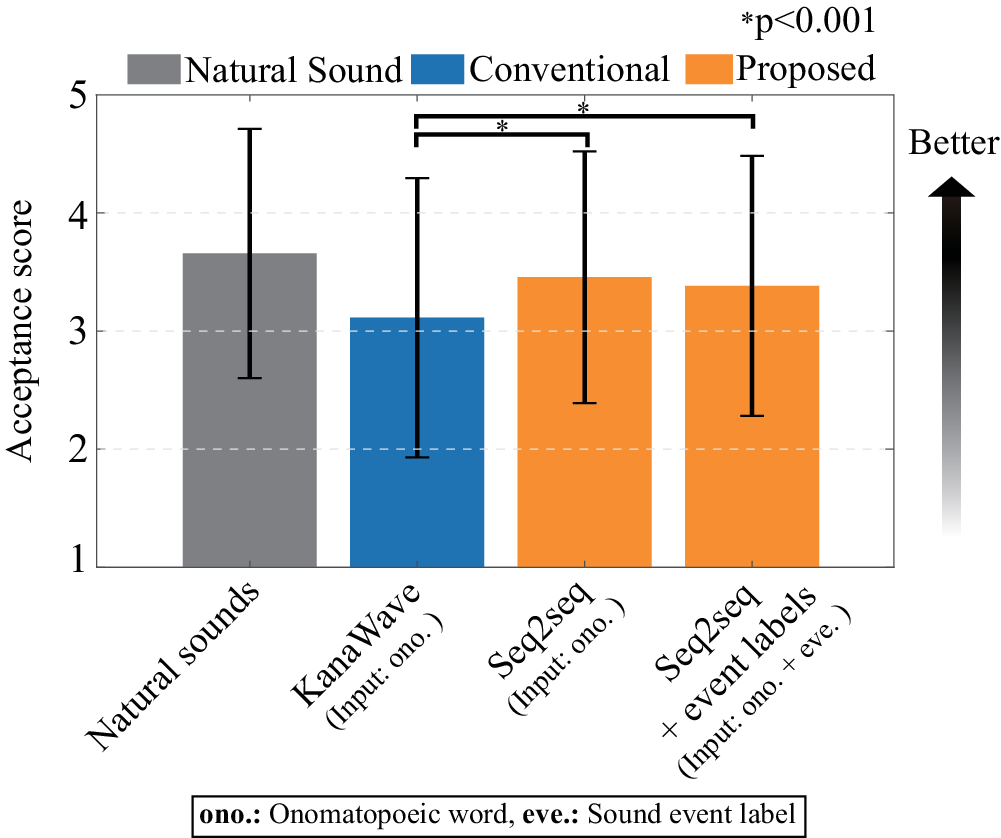}
\vspace{-5pt}
\caption{Acceptance scores of natural and synthesized sounds}
\label{fig:result_acceptamce}
\end{center}
\end{figure}
\subsubsection{Experiment II: evaluation of sound diversity for environmental sounds}
To evaluate diversity for synthesized sounds, we conducted two types of subjective evaluation as follows:
\begin{itemize}
        \item{\bf Experiment II-1: diversity of synthesized sounds for each sound event}\\
        We presented two sounds synthesized by the same method to listeners.
        In the proposed method, sounds are generated using randomly selected onomatopoeic words from the overall dataset as the input in each sound event.
        The listener graded the dissimilarity level between two presented sounds on a scale of 1 (very similar) to 5 (very dissimilar).\\[-5pt]
        \item{\bf Experiment II-2: diversity of synthesized sounds for the same onomatopoeic words}\\
        We presented listeners with a synthesized sound, and the listeners selected the best sound event label that represents the sound from ten choices.
        After listening to each sound synthesized by our methods presented randomly, the listener selected the sound event label that best represented the sound.
\end{itemize}

Each experiment was conducted using a crowdsourcing platform.
Table~\ref{table:experimental_samples} shows the numbers of audio samples and listeners in each experiment.
To compare the synthesis methods, we evaluated the sounds synthesized by the conventional method using WaveNet \cite{okamoto_arXiv_2019} and KanaWave \cite{KanaWave}.
The conventional environmental sound synthesis method using WaveNet utilizes sound event labels as the input to the system to generate sounds.
The conventional method using WaveNet \cite{okamoto_arXiv_2019} does not input onomatopoeic words.
Therefore, we evaluated synthesized sounds by WaveNet in only experiments I-3 and II-1.
KanaWave is the conventional non-statistical method of generating environmental sounds from only onomatopoeic words.
The list of synthesis methods evaluated in each experiment is shown in Table~\ref{table:list_evaluation}.
\begin{figure}[t!]
\centering
\begin{center}
\includegraphics[scale=0.45]{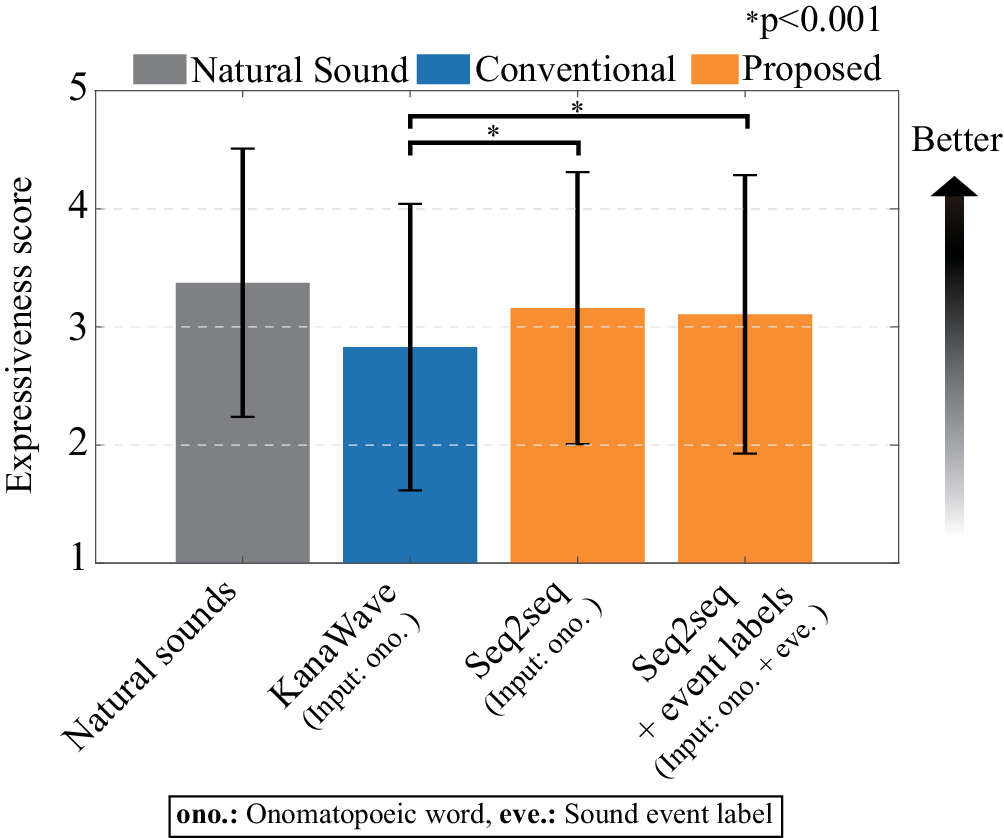}
\vspace{-5pt}
\caption{Expressiveness scores of natural and synthesized sounds}
\label{fig:result_expressiveness}
\end{center}
\vspace{0pt}
\end{figure}
\begin{figure}[t!]
\centering
\begin{center}
\includegraphics[scale=0.30]{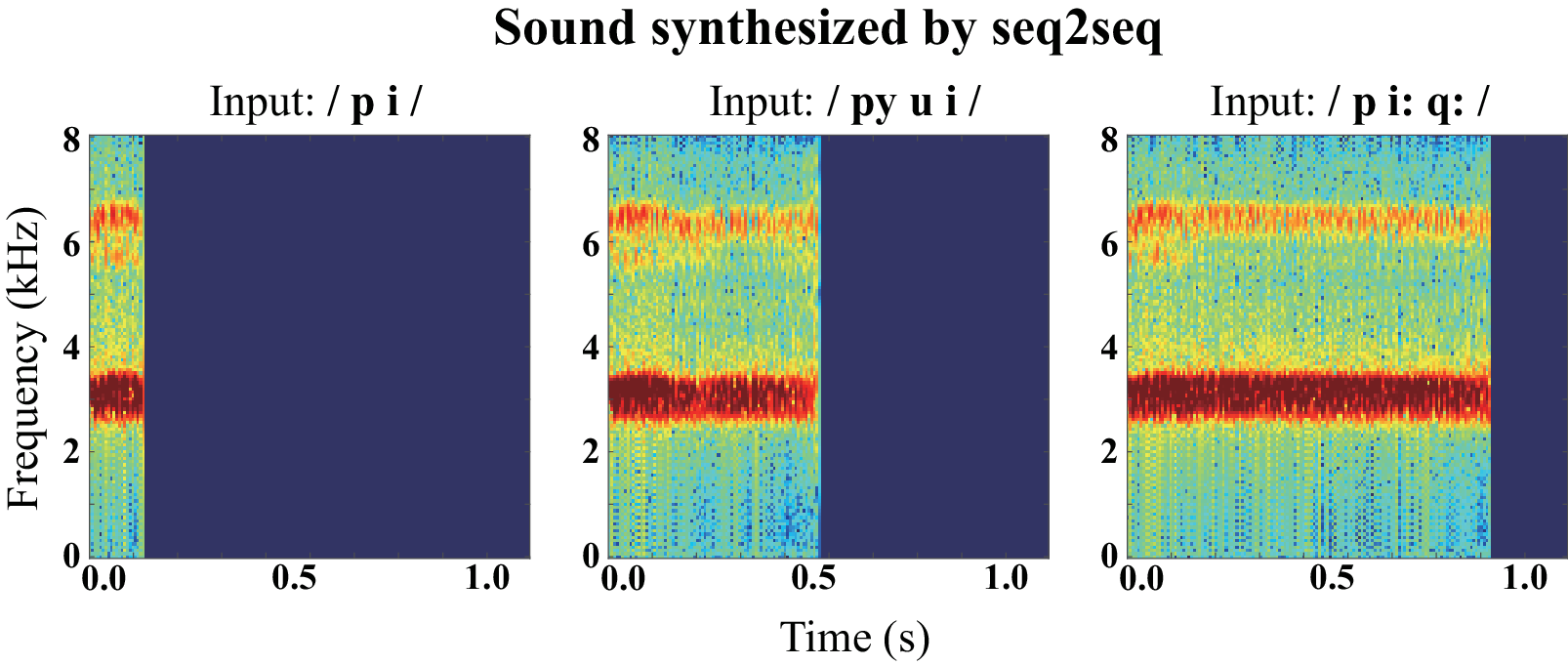}
\vspace{-10pt}
\caption{Spectrograms of environmental sounds synthesized using only onomatopoeic words}
\label{fig:result_whistle_spectrogram}
\end{center}
\end{figure}
%
\subsection{Experimental Results and Discussion}
\subsubsection{Experiment I}
{\bf Experiments I-1 and I-2}: the average acceptance and expressiveness scores of synthesized and natural sounds for onomatopoeic words and their standard deviations are respectively shown in Figs.~\ref{fig:result_acceptamce} and \ref{fig:result_expressiveness}.  
The results show that our proposed methods can generate environmental sounds that are a better representation of onomatopoeic words than those generated by KanaWave.

Figure \ref{fig:result_whistle_spectrogram} shows spectrograms of sounds synthesized by our methods using only onomatopoeic words. 
As shown in Fig.~\ref{fig:result_whistle_spectrogram}, the proposed method can generate diverse environmental sounds.
Also, the longest sound (right) is not the sound given by simply stretching the other sounds (left and center).
Thus, onomatopoeic words are useful for generating diverse sounds with different characteristics, such as sound duration.

\begin{figure}[t!]
\centering
\begin{center}
\includegraphics[scale=0.30]{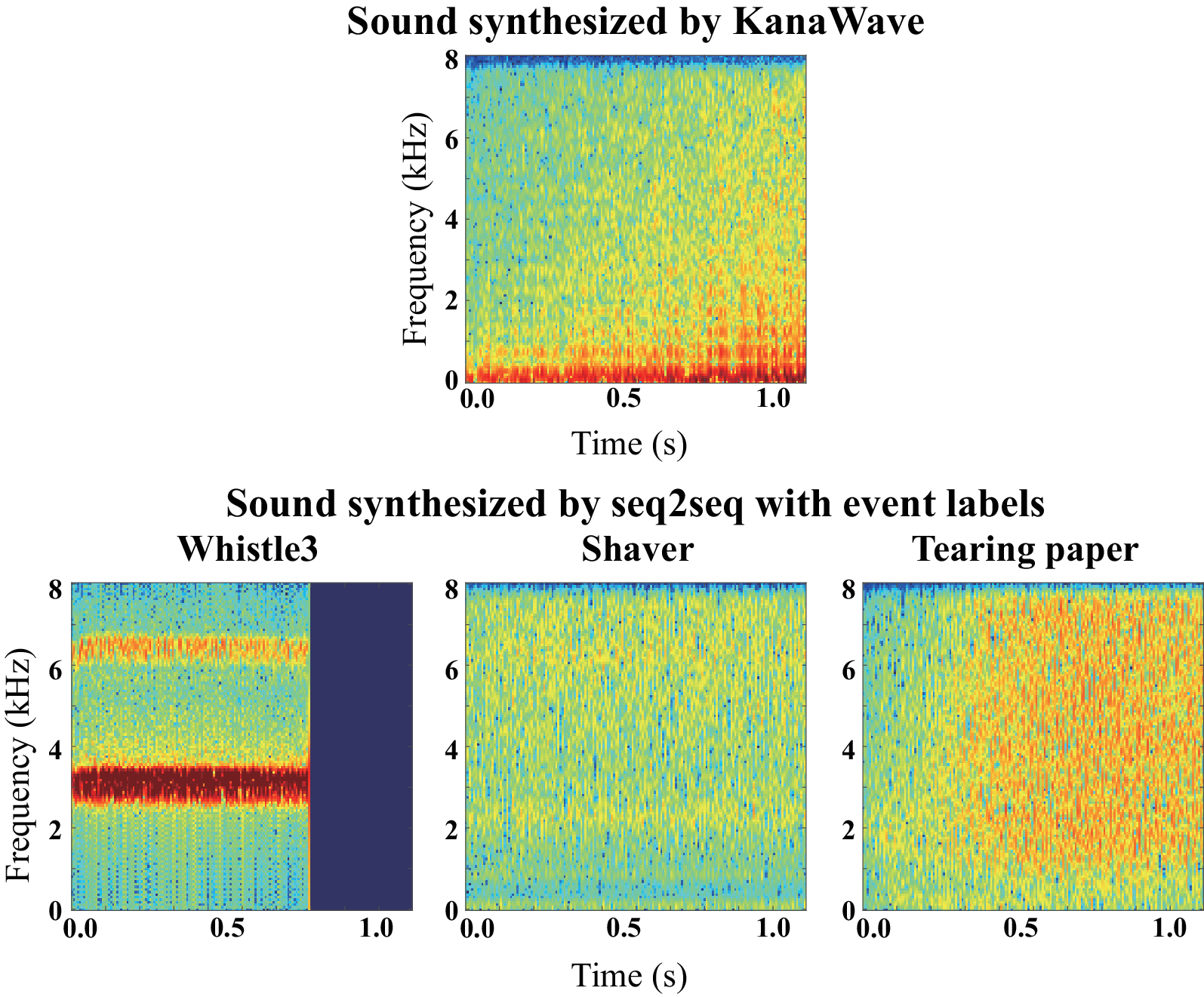}
\vspace{-5pt}
\caption{Spectrograms of environmental sounds synthesized by KanaWave and the proposed method using onomatopoeic words and sound event labels}
\label{fig:result_KanaWave_spectrogram}
\end{center}
\end{figure}

Figure \ref{fig:result_KanaWave_spectrogram} shows the spectrograms of sounds synthesized by KanaWave and the proposed method using both onomatopoeic words and sound event labels.
In Fig.~\ref{fig:result_KanaWave_spectrogram}, each synthesized sound is generated from a phoneme sequence of the onomatopoeic word ``b i i i i i i'' input to the system.
In the proposed method using both onomatopoeic words and sound event labels, we used sound event labels of {\it whistle}, {\it electric shaver}, and {\it tearing paper}.
KanaWave can only generate one type of sound from the same onomatopoeic word.
Therefore, the sound synthesized by KanaWave does not have diversity. 
On the other hand, the proposed method using onomatopoeic words and sound event labels can generate various sounds from the same onomatopoeic word by changing the input sound event labels.

\begin{figure}[t!]
\centering
\begin{center}
\includegraphics[scale=0.45]{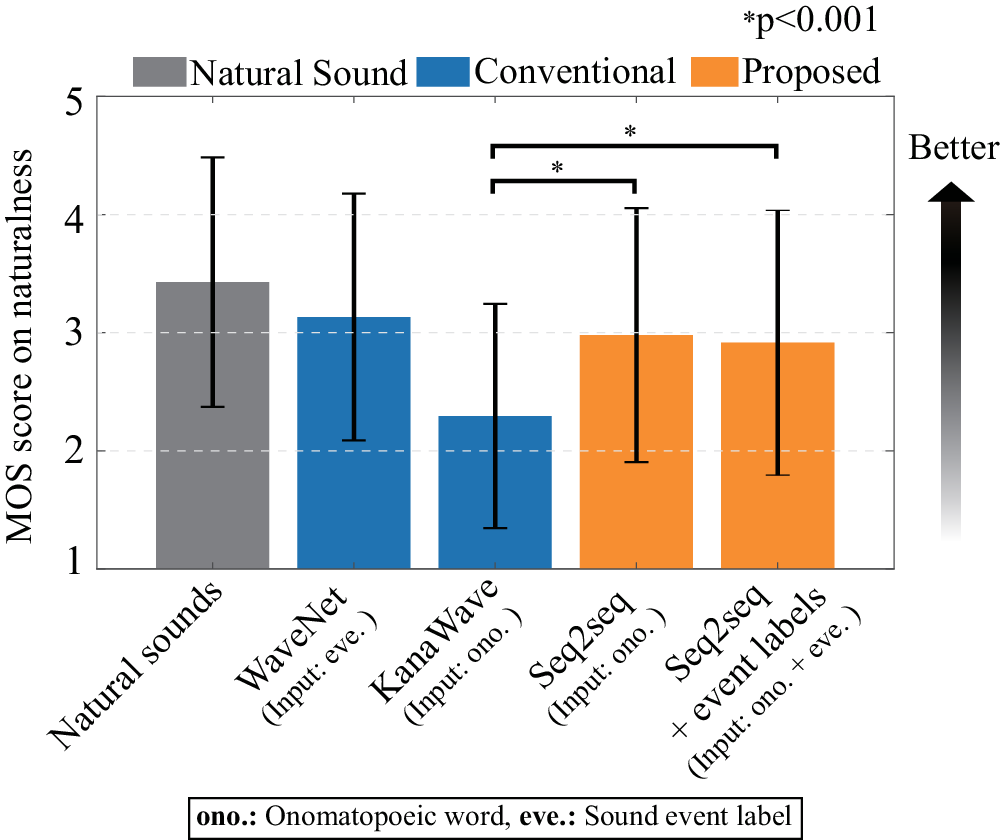}
\vspace{-5pt}
\caption{MOS scores for naturalness of natural and synthesized sounds}
\label{fig:result_naturalness}
\end{center}
\end{figure}

{\bf Experiment I-3}: the average MOS scores for the naturalness of synthesized and natural sounds, and their standard deviations are shown in Fig.~\ref{fig:result_naturalness}.
The results indicate that sounds synthesized by the proposed methods achieve higher naturalness than those synthesized by KanaWave.
The experimental results also show that sounds synthesized by our methods had a similar sound quality to those synthesized by WaveNet. 
Thus, the proposed methods achieve environmental sound synthesis from onomatopoeic words without degrading the sound quality compared with conventional methods. 
In addition, natural sounds still have higher naturalness than sounds synthesized by the proposed methods.
From these results, it is still necessary to develop a method of environmental sound synthesis that can provide quality equivalent to that of natural sounds.
\begin{figure}[t!]
\centering
\begin{center}
\includegraphics[scale=0.44]{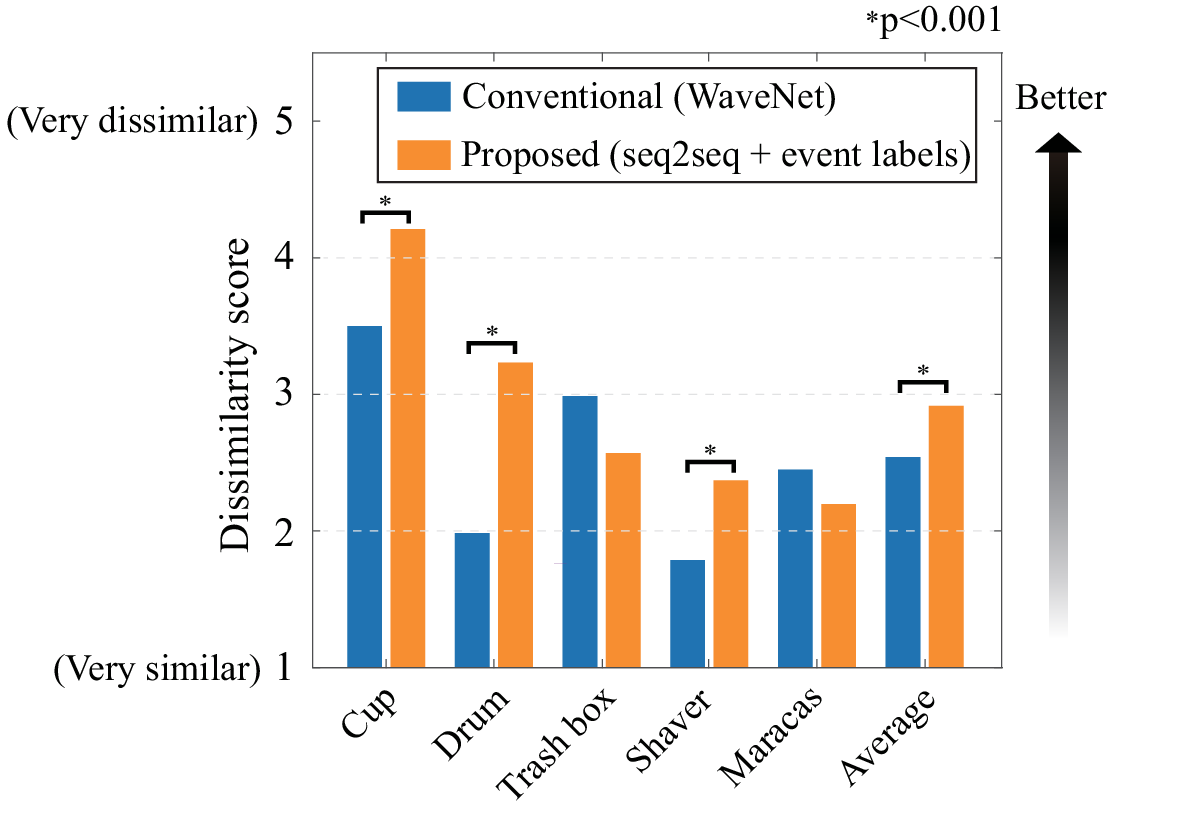}
\vspace{-18pt}
\caption{Dissimilarity scores of synthesized sounds}
\label{fig:result_similarity}
\end{center}
\vspace{0pt}
\end{figure}
\begin{figure*}[t!]
\centering
\begin{center}
\includegraphics[width=\linewidth]{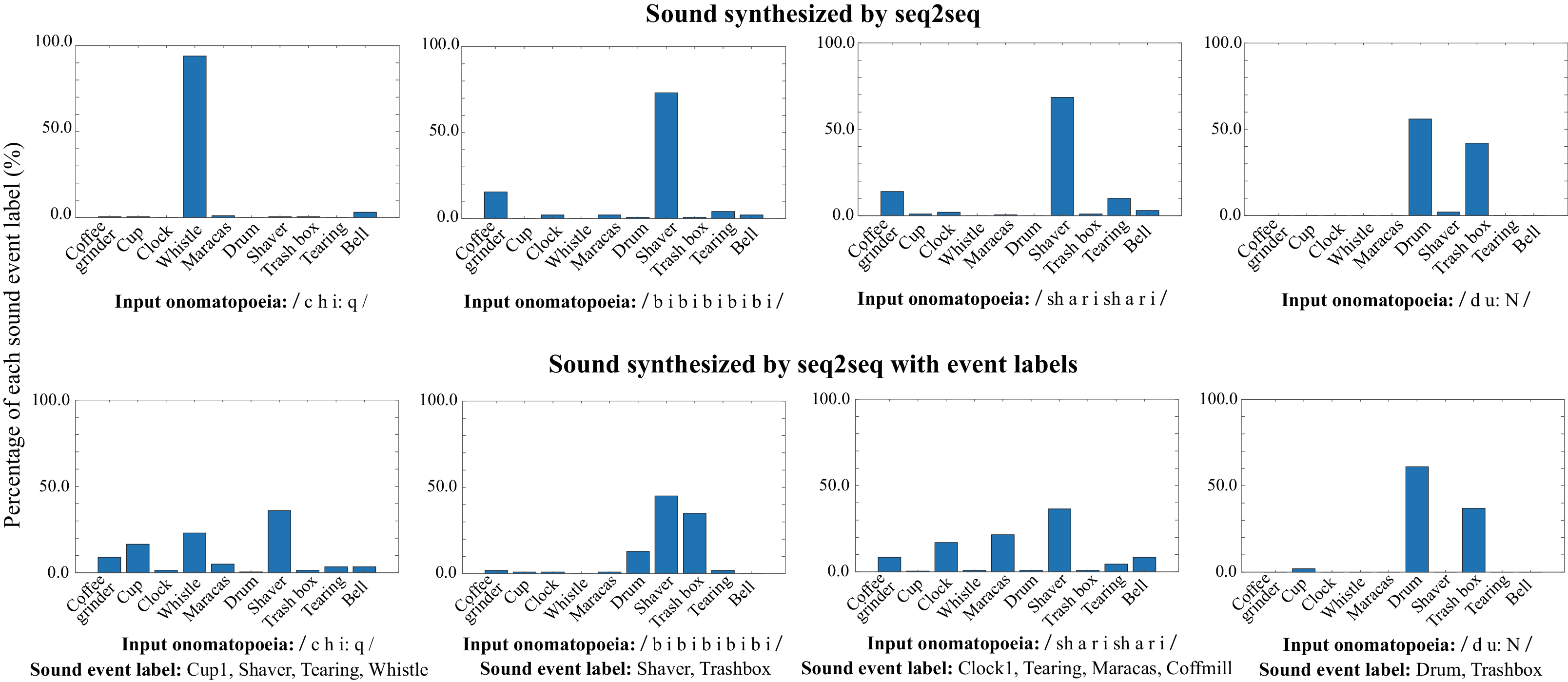}
\vspace{-9pt}
\caption{Number of responses of sound event labels to each sound synthesized by our method using onomatopoeic words}
\label{fig:exp3_hist}
\end{center}
\end{figure*}
%
\subsubsection{Experiment II}
{\bf Experiment II-1}: the average dissimilarity score of the synthesized sound for each sound event is shown in Fig.~\ref{fig:result_similarity}.
In this experiment, 
a high dissimilarity means that there is a rich diversity of synthesized sounds within the same type of event.
The result indicates that the proposed method can generate synthesized sounds with richer diversity than the conventional method using WaveNet.
In particular, the dissimilarity scores of the sound events {\it cup} and {\it drum} synthesized by the proposed methods are high, which indicates that the diversity of the synthesized sounds is richer than that of the sounds obtained by the conventional method.
Thus, the proposed method enables us to generate diverse environmental sounds by using onomatopoeic words. 

{\bf Experiment II-2}: part of the distributions of sound event labels given to the synthesized sound from each onomatopoeic word are shown in Fig.~\ref{fig:exp3_hist}. 
The sounds synthesized by our method using only onomatopoeic words tend to be given only one sound event label. 
On the other hand, the sounds synthesized by our method using both onomatopoeic words and sound event labels tend to be given various sound event labels. 
The entropies of the distribution of a given acoustic event label are 1.70 bit for the method using only onomatopoeic words and 1.82 bit for the method using both onomatopoeic words and sound event labels.
In this experiment, the maximum entropy is 3.02 bit when 10 types of sound event labels equally appear for each synthesized sound.
This result shows that using both onomatopoeic words and sound event labels can represent multiple sound events for the same onomatopoeic word.

\begin{figure}[t!]
\vspace{5pt}
\centering
\includegraphics[scale=0.30]{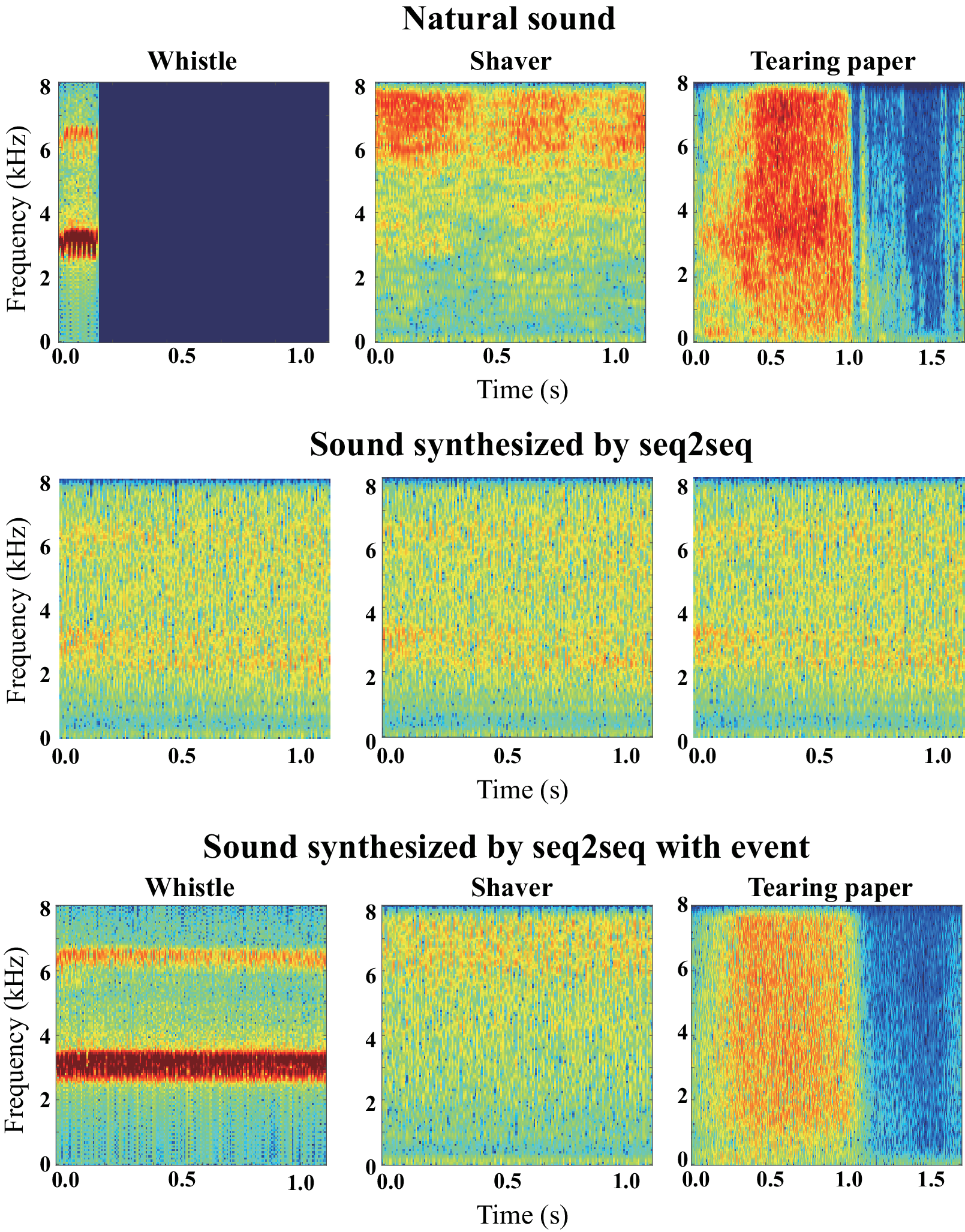}
\vspace{-4pt}
\caption{Spectrograms of synthesized environmental sounds, which are generated from a phoneme sequence of the onomatopoeic word ``b i: iq'', and natural sounds.}
\label{fig:result_spectrogram}
\vspace{-5pt}
\end{figure}

Figure \ref{fig:result_spectrogram} shows spectrograms of natural and synthesized sounds.
In Fig.~\ref{fig:result_spectrogram}, each synthesized sound is generated from a phoneme sequence of the onomatopoeic word ``b i: i q'' input to the system. 
In the proposed method using both onomatopoeic words and sound event labels, we used the sound event labels of {\it whistle}, {\it electric shaver}, and {\it tearing paper}.
As shown in Fig.~\ref{fig:result_spectrogram}, using only onomatopoeic words as an input generates sounds with similar features when the initial values of model parameters in model training are changed.
On the other hand, using both onomatopoeic words and sound event labels, it is possible to generate sounds that capture each sound event's feature depending on the input sound event label. 
These results also show that using sound event labels can control sound events of sound synthesized from onomatopoeic words.
%
\section{Conclusion}
\label{sec:conclusion}
In this paper, we proposed environmental sound synthesis from onomatopoeic words.
We found that the proposed methods can generate sounds with high naturalness and diversity. 
Using sound event labels in addition to onomatopoeic words, we are also able to control not only the time-frequency structure of the synthesized sounds but also the type of sound event.
In the future, we will generate environmental sounds from onomatopoeic words using more types of sound event.
%
\section*{Acknowledgment}
\label{sec:ack}
This  work  was  supported  by  JSPS  KAKENHI  Grant  Number JP19K20304 and ROIS NII Open Collaborative Research 2021 Grant Number 21S0502.

\end{document}